\documentclass[pra,a4paper,nofootinbib,twocolumn,showpacs,preprintnumbers,amsmath,amssymb,floatfix,amstex,superscriptaddress]{revtex4}
\usepackage{graphicx,graphics,wrapfig,rotating}     	% Include figure files
\usepackage{dcolumn}                    		% Align table columns on decimal point
\usepackage{bm,fancybox}                		% bold math
\usepackage{times,euscript,eufrak,oldgerm}              %
\usepackage[german,english]{babel}	

\begin{document}

\title{Decoherence induced by a chaotic enviroment: A quantum walker with a complex coin.}

\author{Leonardo Ermann}
%\email[Email address: ]{ermann@tandar.cnea.gov.ar}
\affiliation{%
Departamento de F\'{\i}sica, Comisi\'{o}n Nacional de Energ\'{\i}a At\'{o}mica.
Avenida del Libertador 8250 (C1429BNP), Buenos Aires, Argentina.
}%
\affiliation{%
Departamento de F\'{\i}sica, FCEyN, UBA, Pabell\'{o}n $1$
Ciudad Universitaria, 1428 Buenos Aires, Argentina.
}%
\author{Juan Pablo Paz}
%\email[Email address: ]{paz@df.uba.ar}
\affiliation{%
Departamento de F\'{\i}sica, FCEyN, UBA, Pabell\'{o}n $1$
Ciudad Universitaria, 1428 Buenos Aires, Argentina.
}%
\affiliation{%
Theoretical Division, LANL, MSB213, Los Alamos, NM 87545,USA}

\author{Marcos Saraceno}
%\email[Email address: ]{saraceno@tandar.cnea.gov.ar}
\affiliation{%
Departamento de F\'{\i}sica, Comisi\'{o}n Nacional de Energ\'{\i}a At\'{o}mica.
Avenida del Libertador 8250 (C1429BNP), Buenos Aires, Argentina.
}%
\affiliation{%
Escuela de Ciencia y Tecnolog\'\i a, Universidad Nacional de San
Mart\'\i n. Alem 3901 (B1653HIM), Villa Ballester, %%Buenos Aires,
Argentina.}
\date{\today}%
\begin{abstract}
We study the differences between the process of decoherence induced by chaotic and regular environments. 
For this we analyze a family of simple models which contain both regular and chaotic environments. In 
all cases the system of interest is a ``quantum walker'', i.e. a quantum particle that can move on a 
lattice with a finite number of sites. The walker interacts with an environment which has a $D$ 
dimensional Hilbert space. The results we obtain suggest that regular and chaotic environments are not 
distinguishable from each other in a (short) timescale $t^*$, which scales with the dimensionality of
the environment as $t^*\propto \log(D)$. However, chaotic environments continue to be effective over
exponentially longer timescales while regular environments tend to reach saturation much sooner. 
We present both numerical and analytical results supporting this conclusion.  The family of chaotic
evolutions we consider includes the so--called quantum multi--baker--map is a particular case. 

\end{abstract}

\maketitle

\section{Introduction}

The study of the transition from quantum to classical physics began with the rise of quantum mechanics
itself \cite{WheelerZurek}. In recent years it became clear that the process of decoherence plays
an esential role in understanding this transition \cite{Decoherence-Reviews}. According to this 
modern view, classicality is an emergent property that is induced on sub--systems due to the 
interaction with their environment. Decoherence is not only important from a fundamental point
of view but also its understanding seems to be crucial to develop new quantum technologies such 
as quantum computation \cite{Chuang}. The role of the environment is esential in the process
of decoherence. In fact, this process can be understood as the consequence of the dynamical 
creation of quantum correlations (entanglement) between the system and its environment. Due
to this process, all quantum imformation initially present in the state of the system is lost
in the correlations with the environment, which effectively measures the state of the system. Due
to this process, the vast majority of the quantum states in the Hilbert space of the system 
become highly unstable. Only  the small subset of states that are relatively immune to the 
interaction with the environment (the so--called pointer states) remain relatively stable. 

In studies of decoherence the environment is usually modelled in a simple way using a phenomenological
approach. In fact, the best known such model is the bosonic bath, where the environment consists of an 
infinite number of harmonic oscillators \cite{Feynman-Vernon,Caldeira-Legget,HPZ1}. Although
it is well known that this model is not universally applicable \cite{Deco-deco} it captures
many of the esential ingredients of the decoherence process and it is quite adequate to 
describe the interaction between quantum systems and large reservoirs which are near some
equilibrium state. Spin baths have been also studied and display some distinctive 
features \cite{Stamp,CPZ,Dobrovitsky}. 

Recently, interest in the study of the effect of the intrinsic complexity of the environment on 
decoherence arose. In fact, there is some evidence that chaotic environments may induce
decoherence more effectively than regular ones \cite{Robin-Zurek}. A particular manifestation
of this higher effectiveness may be the dependence of the decoherence timescale on 
the system-environment coupling strength $\lambda$: regular environments induce a decoherence
rate which is roughly proportional to $\lambda^2$ while unstable \cite{Robin-Zurek} or 
chaotic \cite{Dobrovitsky} environment may display a much weaker dependence with $\lambda$. 
On the other hand, issues such as the heat capacity of a chaotic system as a reservoir
have been addressed \cite{Cohen} and also point at a significant difference betweeen 
the way in which chaotic and regular systems can act as effective reservoirs. 

In this paper we will present a study of the evolution of a quantum system coupled to 
an environment which will be chosen from a family containing both chaotic and regular
representatives. The model we will analyze has recently attracted some attention in the
context of studies of quantum information processing. Thus, we will consider the evolution
of a quantum walker (a quantum particle moving on a finite lattice). The quantum walker
carries a quantum coin which usually consists of a spin $1/2$ particle. The direction of the 
motion of the walker is conditioned on the state of the quantum spin. Here, we will 
consider that the quantum coin is part of a larger quantum system with which it interacts
by means of a unitary operator with either chaotic or regular properties (see below). The 
usual quantum walk has been studied recently as a potentially useful quantum sub--routine 
\cite{Kempe} and the impact of the process of decoherence has also been discussed 
using a variety of tools \cite{Brundeco,Lopez}. 

We will use a family of unitary operators to define the evolution of the environment. This
family was introduced some time ago for a system of qubits \cite{Schackcaves} and contains
a fully integrable member (in such case each qubit evolves independently of the others, 
each of them acting as independent coins \cite{Brun}) and other unitary operators which 
can be seen as the quantization of chaotic systems. The family includes the conventional 
``quantum baker's map'' which is perhaps the simplest and most studied chaotic unitary
map \cite{Voros,Saraceno}. In such case, the complete system we analyze is a variant
of the so--called quantum multibaker chain, which was analyzed before in a different 
context \cite{Wojcik}. 

In our paper we will analyze the behavior of the system (the walker) and show how the interaction
with the environment induces classical behavior on it. We will point out some differences
between the effects induced by the environment when its dynamics is chaotic and regular. Our 
model has a drawback: It does not contain a parameter controlling the strength of the interaction
between the system and the environment. Thus, we cannot detect effects such as the ones
analyzed in \cite{Robin-Zurek}. However, our model will certainly help us to display striking 
differences between regular and chaotic regimes as a function of the dimensionality ($D$) of 
the Hilbert space of the environment. As we will see, regular and chaotic environments 
show some clear diferences in their behavior after relatively short times. 

The paper is organized as follows: In Sec. II we introduce the esential ingredients of the model
we study. We describe the simplest quantum walk on the line and we discuss how it can be coupled 
to a variety of environments whose evolution belongs to the family of the quantum baker maps. 
In Sec. III we show numerical results for the evolution of the system. We analyze first the entropy
induced by the interaction with the environment, which is the magnitude that displays more clearly 
the difference between the chaotic and regular maps. We also analyze the variance of the quantum
walker and a the distance between the phase space representation of the quantum 
walker and their classical counterparts. We present our conclusions in Sec. IV.

\section{The system and the environment}

\subsection{The system: a quantum walker on a ring.}
 
We will consider a quantum walker that moves on a ring. The evolution will be defined by means of  
a sequence of unitary operations (discrete time).  Let $\mathcal{H}_{P}$ be the Hilbert space of the 
walker, which has a finite number of localized states $|j\rangle$ forming a basis that can be denoted 
as $\{|j\rangle ; j=0,\ldots,M-1\}$. The case of an infinite line (i.e. $M\rightarrow\infty$) 
is interesting
and, for initially localized states of the walker, can be obtained from our results for times that
do not exceed $M/2$. If the walker carries a quantum coin consisting of a spin $1/2$ particle, the total 
Hilbert space is $\mathcal{H}=\mathcal{H}_{P}\otimes\mathcal{H}_{C}$ where $\mathcal{H}_{C}$ is the 
space of states of the spin which is spanned by the two states $\{|0\rangle,|1\rangle\}$. 

The evolution of the quantum walker is defined as the succesive application of a unitary 
transformation which is itself built in two steps: 
First, we apply a unitary operator $\left(\hat{I}_P\otimes\hat{C}_C\right)$, which acts non--trivially on 
the coin-space (being the analogue of the classical `coin-flip'). Then, we apply an 
operator that translates 
the state of the walker to the left or to the right depending on the state of the quantum coin. So, the
total evolution in one time--step is defined as 
\begin{equation}
|\Psi(t+1)\rangle=\hat{U}^{\sigma_{z}}\ \left(\hat{I}_P\otimes\hat{C}_C\right)|\Psi(t)\rangle
\end{equation}
where the translation operator $\hat{U}$ acts on the space of the 
walker (as $\hat{U}|j\rangle=|j+1\rangle$)
and $\sigma_{z}$ is the usual Pauli matrix acting in coin space. For the circle 
$\hat{U}$ is diagonal in a basis which is obtained from the position states $|j\rangle$ by means of 
the usual discrete Fourier transfrom. This is the momentum basis defined as 
$|k\rangle=\frac{1}{\sqrt{M}}\sum_{j=0}^{M-1}\exp(-i\frac{2\pi jk}{M})|j\rangle$. It can be easily shown
that $\hat{U}|k\rangle=e^{-i\frac{2\pi}{M}k}|k\rangle$. 
The usual choice for the operator $C_C$, that defines the coin flip, is 
the  so--called Hadamard transformation $H$, whose matrix in the $\{|0\rangle,|1\rangle\}$ 
bases is
\begin{equation}
H=\frac{1}{\sqrt{2}}\left( \begin{array}{cc} 1 & 1\\ 1 & -1\end{array} \right)
\end{equation}\\

In this work we will enlarge the `coin'-space which will consists of $N$ qubits instead of a single one. 
In this case the $D=2^N$--dimensional Hilbert space of the bigger coin will be denoted as 
$\mathcal{H}_{B}$ and the total Hilbert space of the combined walker--coin system is
$\mathcal{H}=\mathcal{H}_{P}\otimes \mathcal{H}_{B}$. At any single instant one qubit (which we 
denote as the ``most significant qubit'' or MSQ) will determine the direction of the motion of the 
walker in the same way as in the ordinary quantum walk. However, we will consider the possibility 
that the evolution of the complex $D$--dimensional coin contains interactions between the different 
qubits. Thus, we can think this model as consisting of an ordinary quantum walk with a spin $1/2$ 
coin which interacts with extra degrees of freedom (in a way that will be specified below). A simple quantum 
circuit describing the evolution is shown in Figure 1. The operator $B_{N,n}$ defines the evolution
of the complex coin and will be described in the next sub--section. 
\begin{figure}[!htp]
\begin{center}
\includegraphics[width=0.42\textwidth]{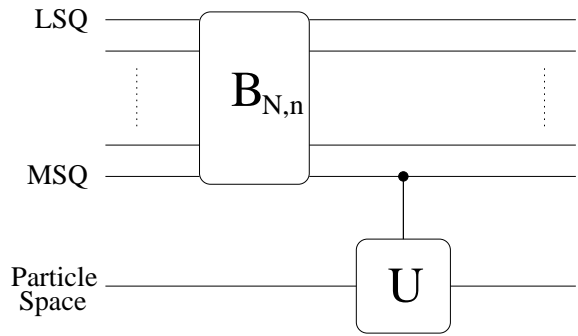}
\end{center}
\caption{\label{fig:QWwithQBM} Circuit representation of the Quantum Walk interacting with families
of Quantum Baker Maps}
\end{figure}

More formally, the evolution of the complete system is 
\begin{equation}
\rho(t)=\hat{M}^t\rho(0)\hat{M}^{\dag t}
\end{equation}
where 
$\hat{M}=\left(\hat{U}\otimes \hat{P}_{0\ MSQ}+\hat{U}^\dagger \otimes \hat{P}_{1\ MSQ}
\right)\left(\hat{I}\otimes\hat{B}_{N,n}\right)$. The operators $P_{0\ MSQ}$ and $P_{1\ MSQ}$ are 
respectively the projectors onto the states $|0\rangle$ and $|1\rangle$ of the 
space of the most significative qubit. As mentioned above, the operator defining the 
evolution on the internal space of the complex (multi--qubit) coin is given by $\hat{B}_{N,n}$
which is described below. 

To study the temporal evolution generated by the operator $\hat{M}$ it is convenient to 
use the momentum basis for the quantum walker. Thus, as the translation operator
$\hat{U}$ is diagonal in such basis we only need to analyze the effect of the operator $M_{k}$, 
wich being defined as $\langle k|M|k'\rangle=\delta_{k,k'}M_{k}$, 
acts in the Hilbert space of the complex coin and has the following matrix form:
\begin{equation}\label{eq:Mk}
\hat{M}_{k}=\left(\begin{array}{cc}e^{-i\varphi_{k}} & 0\\0&
e^{i\varphi_{k}}\end{array}\right)\hat{B}_{N,n}
\end{equation} 
where the first term of the right side is a block--diagonal $D\times D$ matrix and  
$\varphi_{k}=\frac{2\pi k}{M}$.

\subsection{The environment: a family of quantum baker's maps.}

As we mentioned above, our complex coin consists of a set of $N$ qubits. In the $D$--dimensional
Hilbert space we will consider the temporal evolution induced by a family of evolution operators 
which were introduced and studied before \cite{Schackcaves,Cavesscott}. To define these 
operators it is convenient first to introduce the partial Fourier transform $\hat{G}_{n}$
as the operator
\begin{equation}\label{eq:FouParc}
\hat{G}_{n}\equiv\hat{I}_{2^{n}}\otimes\hat{F}_{2^{N-n}}^{\eta,\kappa},\quad\quad
n=0,\ldots,N
\end{equation}
where $\hat{I}_{2^{n}}$ is the identity operator on the first $n$ qubits, and
$\hat{F}_{2^{N-n}}^{\eta,\kappa}$ is the Fourier transform on the remaining qubits. Matrix elements
of this operator are defined (in terms of the so--called Floquet angles $\eta$ and $\kappa$) as
\begin{equation}
\langle k|\hat{F}_{D}^{\eta,\kappa}|j\rangle ={1\over\sqrt{D}} \exp{(-i\frac{2\pi}{D}(j+\eta)(k+\kappa))}.
\end{equation}
%% Borre el circuito para la transformada de Fourier sobre n qubits!!!
%\begin{figure}[htp!]
%\begin{center}
%\includegraphics[width=0.25\textwidth]{FouParc.eps}
%\end{center}
%\caption{\label{fig:FouParc} Circuital representation of the partial Fourier transform acting on the
%N-n less significative qubits.}
%\end{figure}

We define a family of evolution operators which are parametrized by $n$ (the number of qubits 
which are not affected by the partial Fourier transform) and also by 
the Floquet angles $\eta$ and $\kappa$. To simplify the notation the dependence on these two 
parameters will be implicit from here on. The family consists of the 
operators $\hat{B}_{N,n}$ defined as (see \cite{Schackcaves}):
\begin{equation}\label{eq:FamBak}
\hat{B}_{N,n}\equiv
\hat{G}_{n-1}^{-1}\ \hat{S}_{n}\ \hat{G}_{n}
\end{equation}
where the Shift operator $\hat{S}_{n}$ acts only on the first $n$ qubits and is such 
that: 
$\hat{S}_{n}|x_{1}\rangle |x_{2}\rangle \ldots |x_{n}\rangle
 |x_{n+1}\rangle \ldots |x_{N}\rangle=|x_{2}\rangle
 \ldots |x_{n}\rangle |x_{1}\rangle
 |x_{n+1}\rangle \ldots |x_{N}\rangle$.

\begin{figure}[!htp]
\begin{center}
\includegraphics[width=0.4\textwidth]{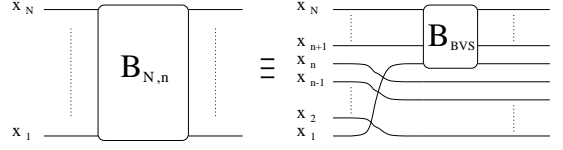}
\end{center}
\caption{\label{fig:cirfambak} Circuit representation of the operator
$\hat{B}_{N,n}$ in terms of the Balasz, Voros and Saraceno baker's map.}
\end{figure}

There is a simpler expression for these operators that can be obtained using the fact that 
the shift $\hat{S}$ commutes with $\hat{G}_{n}$. Then, $\hat{B}_{N,n}$ can be written as
\begin{equation}\label{eq:FamBakCirc}
\hat{B}_{\ N,n}=\left( \hat{I}_{2^{n-1}}\otimes\hat{B}_{\ N-n+1 ,\
1}\right) \circ\hat{S}_{n}.
\end{equation}
Thus, the action of $\hat{B}_{N,n}$ is equivalent to a shift of the $n$ leftmost qubits followed by
application of the map $\hat{B}_{N-n+1,1}$, which acts only on the $N-n+1$ least significant
qubits. The map $\hat{B}_{N-n+1,1}$ is well known in the context of the study of quantum 
chaos. In fact, as the shift $\hat{S}_{1}$ is the identity, we have 
$\hat{B}_{\ N,1}= \hat{F}_{D}^{-1}\ \circ (\hat{I}_{2} \otimes \hat{F}_{D/2})$. Indeed, this map 
was introduced some time ago by Balasz, Voros and Saraceno
as a quantization of the classical baker's map \cite{Voros,Saraceno}. For this reason, it will 
be denoted as $B_{BVS}$. The above equivalence is shown in circuit representation in 
figure \ref{fig:cirfambak}.

On the other hand, it is easy to show that $\hat{B}_{N,N}$, the extreme member of the family (obtained 
when $n=N$) is a map constructed only with swaps and single qubit Fourier transform. Some properties
of this family of operators (such as their entangling power) were studied in \cite{Cavesscott}.
It is interesting to point out that these maps can also be viewed as members of an even larger
family where each member is a product of only two quantized iterations of the classical baker map 
\cite{Leo}. The spectral properties of the maps are interesting. In fact, as will be discussed in 
detail elsewhere \cite{Leo} all the members of the family have rather ``chaotic'' spectra while
the only truly regular member is the extreme case $\hat{B}_{N,N}$ where every qubit evolves 
independently of the rest. 

It is worth commenting on some aspects of the relation between the map $\hat{B}_{N,1}$ and
the quantum version of the classically chaotic baker's map. In fact, the quantization of the 
baker's map can be done on an even-dimensional Hilbert space taking advantadge of some very
simple features of its classical counterpart. Thus, the classical baker's transformation 
acts on a phase space which is the unit square acting on position and momentum coordinates according to
\begin{eqnarray}
q_{i+1}&=&2q_{i}-[2q_{i}]\\
p_{i+1}&=&(p_{i}+[2q_{i}])/2
\end{eqnarray}
where $[q]$ denotes the integer part of $q$.
This map is an example of an intuitive geometrical transformation wich stretches the square by a 
factor of two in the $q$ direction, squeezes by a factor of a half in the $p$ direction, and then 
stacks the right half onto the left. Another advantage of this map is that it has a simple  
symbolic dynamics using the binary {\em Bernoulli shift}. Writing both $q$ and $p$ in 
binary as
$q=0.\epsilon_{0}\epsilon_{1}\ldots=
\sum_{k=0}^{\infty}\epsilon_{k}2^{-k-1}$ and
$p=0.\epsilon_{-1}\epsilon_{-2}\ldots=\sum_{k=1}^{\infty}\epsilon_{-k}2^{-k}$
$(\epsilon_{i}\in{0,1})$, every phase space point can be represented by a bi--infinite 
symbolic string as
\begin{equation}
(p,q)=\ldots\epsilon_{-2}\epsilon_{-1}
\bullet\epsilon_{0} \epsilon_{1}\epsilon_{2}\epsilon_{3}\ldots .
\end{equation}
Then, the action of the baker's map upon symbols turns out to be
\begin{equation}
(p,q)\longrightarrow (p',q')=\ldots\epsilon_{-2}\epsilon_{-1}
\epsilon_{0}\bullet\epsilon_{1}\epsilon_{2}\epsilon_{3}\ldots
\end{equation}
Thus, baker's map is a Bernoulli shift (notice that the most significant bit of the new momentum coordinate
is inherited from the most significant bit of position). 
Using this property, unitary operators that are quantizations of this classical map were
defined \cite{Voros,Saraceno}. The basic idea is to use the unitary operator that maps position
bases onto the momentum bases and let one qubit go through before applying the inverse transformation. 
Thus, the quantum version of baker's map is 
$\hat{B}_{BVS}=\hat{F}_{D}^{-1}\ \circ (\hat{I}_{2} \otimes \hat{F}_{D/2})$. 

It is clear that baker's map can be defined whenever the dimension of the Hilbert space is even.
Moreover, it is well known that although the unitary operator has the spectral properties characterizing
chaotic maps, the case of $D=2^N$ has some peculiar features (where quasi--degeneracies occur??). In the 
coming section we will analyze the properties of an environment with a $D$ dimensional 
Hilbert space in which one of the above operators generate the temporal evolution. In some cases
we will also compare our results with an environment with an even dimensional Hilbert space (which
is not a power of $2$ but is close to one such power). 

\section{Results: regular and chaotic environments.}

We will assume that the initial state of the combined ``walker--coin'' system 
is a tensor product of a localized state for the walker (which from now
on will be denoted simply as ``the particle'') and a pure state of the complex
coin: $|\Psi_{0}\rangle=|0\rangle\otimes|\Phi_{0}\rangle=\sum_1^M\frac{1}{\sqrt{M}}
|k\rangle\otimes |\Phi_{0}\rangle$.
We study the reduced density matrix of the particle obtained by tracing out over 
the coin subspace. 
%We present results which are obtained by averaging over many randomly chosen initial states of the coin. 
The evolution of the probability distibution of the particle is
\begin{equation}
p(x,t)=\frac{1}{M} \sum_{k,k'}\exp{(-i\frac{2\pi}{M} x(k-k'))}
\langle\Phi_{0}|\hat{M}^{t\ \dag}_{k}\hat{M^t}_{k'}|\Phi_{0}\rangle
\end{equation}  
In the case of the classical random walk, $p(x,t)$ has the form of a binomial distribution with a 
width wich spreads as $\sqrt{t}$.

\subsection{Entropy production.}

As the particle and its environment become entangled during the temporal evolution, the reduced
density matrix of the particle losses its purity. A measure of the entaglement between the two
subsystems (particle and coin) is the von Neumann entropy ($S_{V}$) computed from the reduced 
density operators. For simplicity, we will use instead the linear entropy defined as 
$S_{L}= -\log{(Tr[\rho_{P}^{2}])}$ wich is easier to calculate and provides a lower bound to 
$S_{V}$. $S_{L}$ varies between $S_{L}=0$ for pure states and $S_{L}=\ln{D}$ for totally mixed
states (where $D$ is the dimension of the Hilbert space). It is worth mentioning that due to 
the fact that we choose the total state to be pure, the entropy of both subsystems is identical and
is therefore limited by the minimum Hilbert space dimension (which we assume to be given by $D$ as
we are interested in considering the infinite line limit). 

The entropy growth measures the transfer of quantum information from the initial state of the
system onto the quantum correlations with its environment. As mentioned above, at any given instant, 
the entropy measures the number of orthogonal states which are explored in the course of the 
evolution of both
the system and the environment. For this reason, we expect to observe a difference on the entropy 
production
power of chaotic and regular environments. The argument leading to this conclusion may be understood
as follows: Two different localized states of the system can be viewed as generating two 
different effective 
evolutions for the environment. If the evolution is generated by a chaotic unitary map, 
it is known to exhibit extreme sensitivity to perturbations \cite{Peres,Eco}. Then, two different
localized states of the particle will tend to correlate rapidly with approximately orthogonal states
of the environment. Then, the entropy will grow until all available orthogonal directions in Hilbert
space are explored. Therefore, for chaotic environments one expects the entropy to saturate at 
levels which are of the order of $\log(D)$. For regular environments one expects to be in the 
opposite regime: the evolution will tend to explore a number of dimensions which should be much 
smaller than in the chaotic case. 

\begin{figure}[!htp]
\begin{center}
\includegraphics[width=0.5\textwidth]{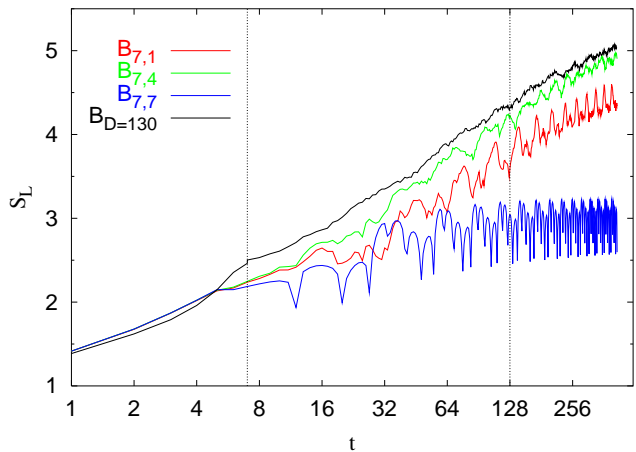}
\end{center}
\caption{\label{fig:SlinQ8Flo05}  
Linear entropy as a function of time for an enviroment of seven qubits evolving with 
some typical unitary operators of the family $B_{7,n}$. Floquet angles are fixed at
$\eta=\kappa=0.5$. The initial state of the particle is localized in position and 
the initial state of the enviroment was chosen as $\frac{1}{\sqrt{2}}(|0\rangle+i|1\rangle)$ 
for each qubit. Results are shown for a similar initial state for an environment with a 
$D=130$ dimensional Hilbert space evolving with the baker's map (black line).}
\end{figure}

The time dependence of the linear entropy $S_{L}$ is displayed in Figure 
\ref{fig:SlinQ8Flo05} for some representative members of the family of environmental 
evolutions $B_{7,n}$ (we show the results corresponding to $\eta=\kappa=0.5$, but the 
behavior is qualitatively similar for other Floquet angles). 

It is clear that a very different behavior is observed for the regular member of the family (the 
map $B_{N,N}$). 
In such case the entropy production saturates at a level which is of the order of $S_0=\log(\log(D))$ (as 
$D=128$ this value is close to $S_0\approx 2.8$, see below). This behavior is also seen to be 
independent of the initial condition. As mentioned above, this can be understood as a consequence
of the small generation of entanglement between the qubits of the environment. 
On the other hand, all the other members of the baker's family $B_{N,n}$ for $n=1,\ldots,N-1$ have a 
similar behavior. The entropy continue growing approaching an asymptotic value which is 
of the order of a fraction of $\log(D)$. Entropy continues growing for  times which scale proportionally
to $D$, the Hilbert space dimensionality. It is worth mentioning that within the family of maps 
$B_{N,n}$ the ones that achieve maximal entropy growth correspond to intermediate values of 
$n$, in agreement with the results obtained in \cite{Cavesscott}. The fact that the maximal 
value of $\log(D)$ is not attained can be attributed to the quasi--degeneracies present
in the spectrum of the baker's map for dimensions which are a power of two. In fact, in Figure
\ref{fig:SlinQ8Flo05} we also show the entropy production from a chaotic environment whose 
Hilbert space dimension is $D=130$ (which is an even number close to a power of two). It is clear
that the entropy for this map is larger than the rest. This supports the argument stating that
an environment that is more chaotic is able to generate more entropy. It is also consistent with
the claims of \cite{Cavesscott} concerning the fact that spatial symmetries in the quantum baker's 
map are responsible for deviations from the predictions of random matrix theory. 

\begin{figure}[!htp]
\begin{center}
\includegraphics[width=0.45\textwidth]{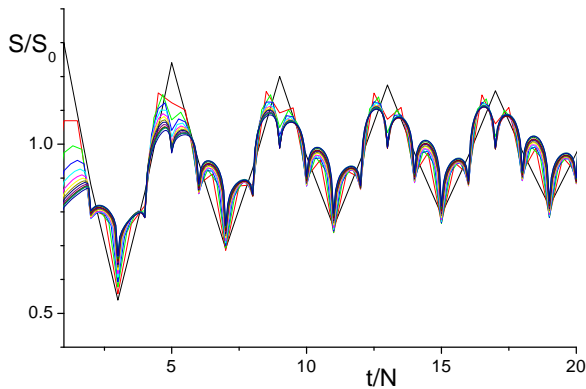}
\end{center}
\caption{\label{fig:SlinMC}  
Linear entropy normalized vs $t/N$ for many coins walk, $B_{N,N}$ and $\eta=\kappa=0$, up to $N=8$. 
The particle was localized in position and the initial state of the enviroment is
$\frac{1}{\sqrt{2}}(|0\rangle+i|1\rangle)$ for each qubit}
\end{figure}

The behavior of the regular environment $\hat{B}_{N,N}$ can be examined using analytic tools. In 
fact, we can show that after the Eherenfest time $\log(D)$ the linear entropy $S_{L}$ oscillates 
around the saturation value $S_0$ with period which is identical to the number of qubits $N$.
In fact, we can obtain a universal curve for the normalized linear entropy ($S_{L}/S_{0}$) as a 
function of the rescaled time $t/N$. This is shown in Figure \ref{fig:SlinMC}. 

It is also possible to obtain a good estimate for the saturation value of the linear entropy. This
is shown in Figure \ref{fig:S0MCF05} where the behavior of $S_0$ (the saturation value of $S_L$)
as a function of the number of qubits $N$ is displayed. This saturation value 
is bounded by $\log(N)$ (which in turn implies that the linear
entropy for regular environment is bounded by $\log(\log(D))$. 
\begin{figure}[!htp]
\begin{center}
\includegraphics[width=0.45\textwidth]{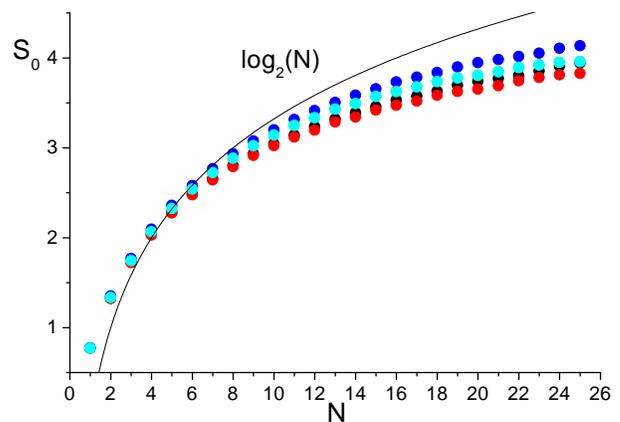}
\end{center}
\caption{\label{fig:S0MCF05.eps}  
(Color on--line) 
Saturation value of linear entropy for many coins walk vs. number of coins. The initial state 
for each coin is
$|\psi_{0}\rangle=|0\rangle$ (black), $|\psi_{0}\rangle=\frac{1}{\sqrt{2}}(|0\rangle+e^{i\frac{3\pi}{4}}|1\rangle)$ (red), 
$|\psi_{0}\rangle=\frac{1}{\sqrt{2}}(|0\rangle+i|1\rangle)$ (blue) with $\eta=\kappa=0$.
As argued in the text, the $\log(\log(D))$ curve establishes an upper bound for the 
saturation.}
\end{figure}
In the above discussion we referred to the many--coin map as a regular system. The reason
for our use of this terminology is the following: As the coins do not interact the spectrum
of the evolution operatir is highly degenerate. It is worth mentioning that this is the 
only sense in which this can be viewed as an integrable system since it does not have a 
classical analogue. 

\subsection{Quantum and classical behavior of the spread of the wave--packet.}

\begin{figure}[!htp]
\begin{center}
\includegraphics[width=0.45\textwidth]{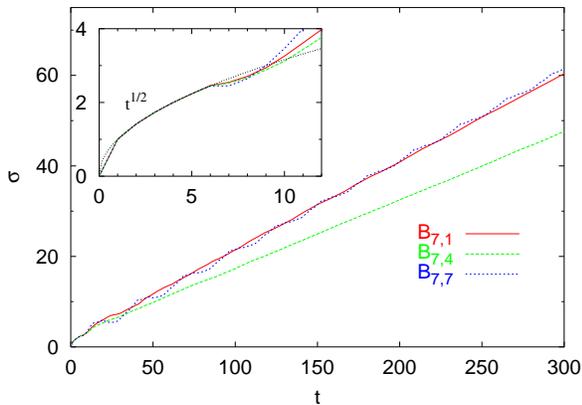}
\end{center}
\caption{\label{fig:VarQ8Flo05} Standard deviation as a function of time for the enviroment of 7qubits ($B_{7,n}$ with
$n=1,\ldots,7$) and $\eta=\kappa=0.5$ in logarithmic scale. The
particle was localized in position and the initial state of the enviroment is
$|\phi_{0}\rangle=\frac{1}{\sqrt{2}}(|0\rangle+i|1\rangle)$ for each qubit}
\end{figure}

The study of the variance of the particle's position, that can be defined as  
$\sigma^{2}=\langle x^{2} \rangle - \langle x\rangle^{2}$ can be useful to signal the 
transition from a classical to a quantum regime. From the above study of the entropy we 
expect that both chaotic and regular systems should be quite efficient to enforce classical 
behavior for times which are of the order of $\log(D)$ (the Ehrenfest time). 
For larger times one expects regular environments to loose its ability to induce classicality. 
Thus, for larger times one expects the particle to spread according to the quantum predictions
while for shorter times it should behave classically (although at first sight this may sound
counter--intuitive, for this system one really expects to see a classical--to--quantum transition!).
For the classical random walk, it is well known that the variance grows diffusively (i.e., 
linearly with time). In turn, for the ordinary quantum walk (with no decoherence mechanism)
the variance grows quadratically with time. 
In figure \ref{fig:VarQ8Flo05} we show the standard deviation ($\sigma$) as a function of time for some representative
members of the $B_{7,n}$ family (again, we display results for $\eta=\kappa=0.5$ and for an 
initial state of the complex coin which is a tensor product of 
$\frac{1}{\sqrt{2}}(|0\rangle+i|1\rangle)$ for each qubit).  

\begin{figure}[!htp]
\begin{center}
\includegraphics[width=0.5\textwidth]{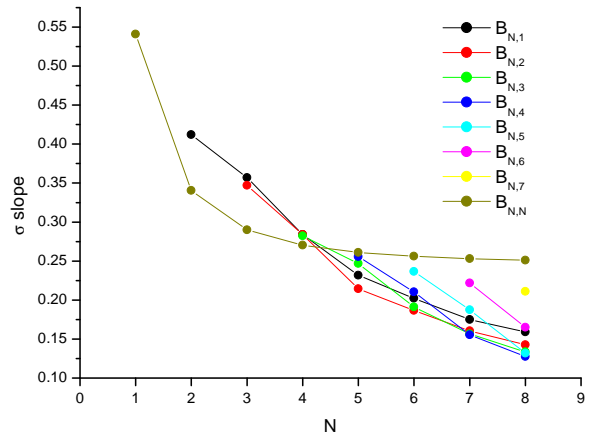}
\end{center}
\caption{\label{fig:PendVar} The time derivative of the standard deviation $\sigma$ for long times 
as a function of the number of qubits in the environment. For regular dynamics the slope approaches
a constant while for chaotic ones it decays. Thus, the position variance grows slower than linear
for long periods of time, which is evidence in favor of the higher efficiency of chaotic environments.}
\end{figure}

As expected, the standard deviation (SD) grows diffusively for short periods of time both for regular and 
chaotic environments. This is seen in the inset of Figure \ref{fig:VarQ8Flo05} where no noticeable
difference between chaotic and regular environments arise before the Ehrenfest time. 
For larger times the evolution is more complex. For the regular environment the growth
is clearly linear signalling a transition from classical to quantum, as expected. The
behavior for chaotic evolutions is harder to visualize. At first glance the behavior of the
SD seems to be linear with time. However, there is a clear separation between the slope
of the line which is attained for the regular case and for the chaotic one being substantially
smaller for the latter. Moreover we observe that by enlarging the dimensionality of the 
environment the slope of the SD for the chaotic environment decreases (while it remains
constant for the regular case). The behavior of the slope (the time derivative of the SD)
is displayed in Figure \ref{fig:PendVar}. The conclusion is that for large chaotic environments 
the time derivative of the variance tends to very small values as $D$ increases. Therefore, 
the growth of the variance will be slower than linear, which is a manifestation of their larger
efficiency as compared with regular ones. 

The behavior of the variance for the regular map can be understood by generalizing some 
of the results obtained in  \cite{Brun} to include arbitrary Floquet angles in the Fourier transform. 
Then, one can show that the long time behavior of the variance is (for $|0\rangle$ position as initial state with 
$\eta=\kappa=0$)
\begin{equation}
\sigma^{2}(t)=\frac{3-2\sqrt{2}+1/N}{4\sqrt{2}}t^{2}+O(t)+\text{(oscillatory terms)}
\end{equation}
where N is the number of coins. One can show that changing Floquet angles is equivalent to 
changing the initial coin state \cite{bach}. Using this we obtained results which show that for
long times the time derivative of the variance approaches a constant value for large number 
of qubits. 

\subsection{Approach to classical phase space distributions.}

\begin{figure}[!htp]
\begin{center}
\includegraphics[width=0.45\textwidth]{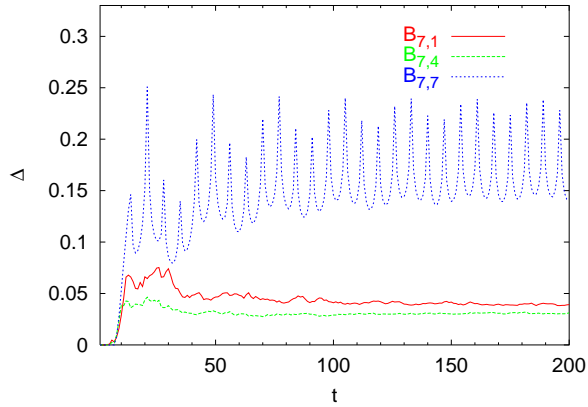}
\end{center}
\caption{\label{fig:DistEFQ8Flo05}  
Evolution of phase-space  distance  ($\delta$) for $B_{7,n}$ with $n=1,\ldots,7$ and $\eta=\kappa=0.5$.
The particle was localized in position and the initial state of the enviroment is $\frac{1}{\sqrt{2}}
(|0\rangle+i|1\rangle)$ for each qubit}
\end{figure}

Another interesting aspect of the quantum to classical transition is the study of the way in which
quantum phase space quasi--distributions (like Wigner functions \cite{Scully}) 
approach their classical counterparts \cite{PHZ}. To study this we use the discrete version 
of the Wigner function \cite{Miquel}. For a system with an $M$-dimmensional Hilbert space the discrete
Wigner function can be defined in a phase space grid of $2M\times 2M$ points. Thus, the Wigner
function is the expectation value of the so--called phase-space point operators which are defined as 
$A(q,p)=U^{q}RV^{-p}\exp{(i\pi pq/M)}$. Here $U$ and $V$ are the cyclic shift operator in position
and momentum respectively ($U|n\rangle=|n+1\rangle$ and $V|k\rangle=|k+1\rangle$), and $R$ is the reflection
operator (wich in the position basis act as $R|n\rangle=|-n\rangle$). Phase-space operators are unitary,
Hermitian and form a complete orthogonal basis of the space operators. As mentioned above, the Wigner 
function is defined as $W(q,p)=Tr[\rho A(q,p)]/M$. This function not only provides a complete 
description of the quantum state but also can be used to compute marginal probability distributions
by adding its values along arbitrary phase space lines (see \cite{Miquel}). 
To study how fast the quantum state approaches a classical distribution we define a distance between 
two such distributions as $\delta_{1,2}\equiv \sum_{q,p} (W_{1}(q,p)-W_{2}(q,p))^{2}$. 
We analyze the distance between the Wigner function at any given instant and the classical distribution
corresponding to the classical random walk. The behavior of this measure is displayed in 
Figure \ref{fig:DistEFQ8Flo05} for some representative members of the  $B_{7,n}$ family. It can be
seen that the regular map ($B_{7,7}$) significatively differs with respect to the chaotic maps. 
Again the most decoherence is attained by the chaotic environment. While interacting with the 
regular environment, the quantum state of the system looses track of the classical state after a 
short time. These results are in agreement with the ones obtained for the entropy and the position 
variance. 

\section{Conclusions}

We studied a model where the decoherence induced on a system by its interaction with an environment
can be analyzed both for an environment endowed with a regular or a chaotic evolution. As the Hilbert 
space of the environment has a finite dimension $D$, the system cannot display a truly dissipative 
behavior. In fact, after a finite time the environment ceases to be effective. For this reason, after
this time quantum effects on the system can be recovered. 
Our results provide a strong evidence showing that a chaotic environment can be efficient over much
longer timescales than regular ones. In fact, the time over which a chaotic environment is effective
seems to scale as a power of the Hilbert space dimension $D$. On the other hand, a regular environment
is effective only for a much shorter timescale, which is of the order of the 
Eherenfest time $\log(D)$. For such short timescales both environment are truly indistinguishable 
from each other.

%\bibliographystyle{plain}

%\bibliography{paper-refsl}

\begin{thebibliography}{99}

\bibitem{WheelerZurek} see {\it Quantum theory and measurement}, edited by J.A. Wheeler and W.H. Zurek, Princeton Univ. Press
(1983). 

\bibitem{Decoherence-Reviews} for a review see J. P. Paz and W. H. Zurek,
in "Coherent matter waves, Les Houches Session LXXII", edited by
R Kaiser, C Westbrook and F David, EDP Sciences, Springer Verlag
(Berlin) (2001) 533-614; W. Zurek, {\it Rev. Mod. Phys.} {\bf 75}, 715 (2003).

\bibitem{Chuang} A. Nielsen y I. Chuang, \emph{Quantum Computation and Quantum
Information}, Cambridge University Press (2000).

\bibitem{Feynman-Vernon} R.P. Feynman and F.L. Vernon, {\it Ann. Phys.} {\bf 24},  118 (1963). 

\bibitem{Caldeira-Legget} A.O. Caldeira and A.J. Leggett, {\it Physica} 
{\bf 121A}, 587-616 (1983); {\it Phys. Rev.} {\bf A 31}, 1059 (1985).

\bibitem{HPZ1} B.L. Hu, J.P. Paz and Y. Zhang, Y., {\it Phys. Rev.} {\bf D 45}, 2843 (1992).

\bibitem{Deco-deco} J.R. Anglin, J.P. Paz and W.H. Zurek,  
{\it Phys. Rev.} {\bf A 53}, 4041 (1997).

\bibitem{Stamp} N.V. Prokof'ev and P.C.M.. Stamp, Rep. Prog. Phys. {\bf 63}, 669 (2000).

\bibitem{Dobrovitsky} V.V. Dobrovitsky and H.A. De Raedt, Phys. Rev. {\bf E 67} 056702 (2003).

\bibitem{CPZ} F. Cucchietti, J.P. Paz and W.H. Zurek, {\it Decoherence from a spin environment}, e--print
quant-ph/0508xxx.

\bibitem{Robin-Zurek} R. Blume--Kohout and W.H. Zurek, Phys. Rev. {\bf A 68}, 032104 (2003).

\bibitem{Cohen} D. Cohen and T. Kottos, Phys. Rev. {\bf E 69}, 55201 (2004).

\bibitem{Kempe} J. Kempe, Contemporary Physics \textbf{44}, 307-327 (2003), e-print
quant-ph/0303081.

\bibitem{Brundeco} T.A. Brun, H.A. Carteret and A. Ambainis, Phys. Rev. A \textbf{67}, 032304 (2003).


\bibitem{Lopez} C.C. L\'{o}pez and J.P. Paz , Phys. Rev. A \textbf{68}, 052305 (2003).

\bibitem{Schackcaves} 
R. Shack and M.C. Caves, Applicable Algebra in Engineering, Communication and Computing, 
{\bf $10$}, $305$ (2000).

\bibitem{Brun} T.A. Brun, H.A. Carteret and A. Ambainis, Phys. Rev. A \textbf{67}, 052317 (2003).

\bibitem{Voros} N.L. Balazs and A. Voros, Ann. Phys, \textbf{190} (1989) 1.

\bibitem{Saraceno} M. Saraceno , Ann. Phys., \textbf{199} (1990) 37.

\bibitem{Wojcik} D.K. W\'{o}jcik and J.R. Dorfman, Physica D, \textbf{187}, 223-243 (2004).


\bibitem{Cavesscott} A.J. Scott y M.C. Caves, J. Phys. A {\bf36} 9553 (2003), quant-ph/0305046 (2003).

\bibitem{Leo} L. Ermann and M. Saraceno in preparation.

\bibitem{Peres} {\it ``Quantum theory concepts and methods''}, A. Peres, Kluwer Univ. Press (1994); see Chapter 12. 

\bibitem{Eco} H. Pastawski, G.  Usaj, and P. Levstein, {\it Chem. Phys. Lett.} {\bf 261} 329 (1996);
R. Jalabert and H. Pastawski, Phys. Rev. Lett. (2001); F. Cucchietti, D. 
Dalvit, J.P. Paz and W.H. Zurek, Phys. Rev. Lett. {\bf 91} 210403 (2003). 

\bibitem{Scully} M. Scully, M. Hillery and E. Wigner, Phys. Rep. \textbf{106}, 121 (1984).

\bibitem{PHZ} J.P. Paz, S.  Habib  and W.H. Zurek, 
{\it Phys. Rev.} {\bf D 47}, 488 (1993).

\bibitem{Miquel} C. Miquel, J.P. Paz, M. Saraceno, Phys. Rev. A \textbf{65}, 062309 (2002).

\bibitem{bach} E. Bach, S. Coppersmith, M.P. Goldschen, R. Joynt, J. Watrous (2002) e-print
quant-ph/0207008.

\end{thebibliography}

\end{document}